\documentclass[
  prl,
  reprint,
  superscriptaddress,
  amsmath,
  amssymb,
  floatfix
]{revtex4-2}

\usepackage[T1]{fontenc}
\usepackage[utf8]{inputenc}
\usepackage{graphicx}
\usepackage{dcolumn}
\usepackage{bm}
\usepackage{physics}
\usepackage{hyperref}
\usepackage{xcolor}
\hypersetup{
  colorlinks=true,
  linkcolor=blue,
  citecolor=blue,
  urlcolor=blue
}

\begin{document}

\title{\makebox[\textwidth][c]{Molecular Electrodynamic Relaxation in Dielectric and Plasmonic Cavities}}

\author{Gang Meng}
\affiliation{Department of Physics and Astronomy, University of Delaware, Newark, Delaware 19716, USA}

\author{Tao E. Li}
\email{taoeli@udel.edu}
\affiliation{Department of Physics and Astronomy, University of Delaware, Newark, Delaware 19716, USA}

\date{\today}

\begin{abstract}
Molecular radiative emission is a fundamental relaxation process that governs excited-state lifetimes and significantly influences photophysical dynamics, but direct self-consistent Maxwell--Schr\"odinger simulations of molecular radiation suffer from numerical instabilities in evaluating radiative self-interaction.   Herein, this issue is resolved within the self-consistent propagation of classical Maxwell's equations coupled to real-time density-functional electronic dynamics and their tight-binding approximation. By carefully removing the longitudinal components of radiative self-reaction, this framework quantitatively reproduces radiative lifetimes of atomic and molecular emitters in vacuum and further captures cavity-modified emission at weak and strong coupling regimes beyond perturbation theory. For molecules weakly coupled to an optical microcavity, the cavity Purcell effect is shown to reach maximum using a modestly red-detuned cavity due to nuclear thermal motion. Moreover, for a single methylene blue molecule strongly coupled to a realistic nanoparticle-on-mirror plasmonic cavity, the predicted molecular electrodynamical relaxation  is accelerated to approximately 11 femtoseconds, directly competing with nonadiabatic nuclear dynamics. This work establishes a route to  first-principles design of material radiative properties in a wide range of dielectric environments.
\end{abstract}

\maketitle

\textit{Introduction}---Following optical excitation, electronically excited molecules and materials can relax through two classes of pathways: nonradiative and radiative decay. In nonradiative relaxation, electronic energy is transferred to nuclear degrees of freedom through electron-nuclear nonadiabatic coupling, giving rise to processes such as photochemical reactions and internal conversion \cite{Yarkony2012, Crespo-Otero2018a}. Alternatively, electronic energy can also radiatively dissipate into surrounding electromagnetic (EM) environments through photon emission. Nonadiabatic relaxation has received extensive attention over radiative processes in chemical physics, partly because of the distinct timescales of two processes. After all, nonadiabatic relaxation usually occurs on femtosecond-to-picosecond timescales, whereas molecular radiative decay commonly occurs on timescales of nanoseconds or longer \cite{Berezin2010, Nelson2020}.

However, this timescale separation is challenged by an increasing number of experiments. A recent example is the use of polariton-enhanced Purcell effects in blue phosphorescent organic light-emitting diodes (OLEDs), where the accelerated radiative decay rate reduces the destructive triplet density, thus extending the device operational lifetime \cite{zhao2024stable}. As another example, in a plasmonic nanoparticle-on-mirror (NPoM) cavity, methylene blue molecules have been shown to strongly couple  to plasmons confined in the nanocavity, producing hybrid plasmon-exciton polariton states \cite{Chikkaraddy2016}. Such a nanocavity is also estimated to accelerate molecular radiative decay by approximately six orders of magnitude \cite{Chikkaraddy2016}, a timescale which can directly compete with nonadiabatic nuclear dynamics. It thus becomes practically important for first-principles, non-perturbative prediction of molecular spontaneous emission in complex dielectric environments \cite{Torma2015, ebbesen2016hybrid, akselrod2014probing, rose2014control,Li2018Spontaneous,Chen2018Spontaneous,Hoffmann2019,Hoffmann2019Benchmark,Bustamante2021,schafer2021TDDFT,Hirai2023}.

A computationally appealing approach for attacking this problem is the self-consistent propagation of coupled Maxwell--Schr\"odinger equations \cite{Chen2010, Ruggenthaler2011, Li2016, Trivedi2017, Yamada2018, Li2018Spontaneous, Li2019PRA, Vargas2024, bustamante2025molecular, Bonafe2025,Ziolkowski1995, Neuhauser2007Nanopolaritonics, Lopata2009, Sukharev2011,Gao2012, Yabana2012Multiscale, Sakko2014Hybrid, Jestadt2019EMPKS, Noda2019SALMON, TancogneDejean2020Octopus}.
Nevertheless, this semiclassical scheme was long hampered by severe numerical instabilities and spurious frequency shifts arising from the singular radiative self-reaction between molecules and their emitted classical electromagnetic (EM) fields \cite{Lopata2009,Deinega2014,Schelew2017}, not to mention the length-scale mismatch between molecular emitters and dielectric media as well as the time-scale mismatch between radiative emission and electronic dynamics \cite{Fregoni2022, weight2026ab}. 

For first-principles description of spontaneous emission of real atomic systems in vacuum,  Scherlis and coworkers bypassed this numerical instability via an augmented real-time time-dependent density functional theory (RT-TDDFT) approach \cite{Bustamante2021}. In this method, an Abraham--Lorentz-type one-electron term is added to the electronic Kohn--Sham matrix  without dealing with explicit EM environments. A complementary strategy is further proposed by Sch{\"a}fer and Johansson  for incorporating radiative-reaction correction within RT-TDDFT indirectly via Green tensors \cite{schafer2021TDDFT}. As these approaches require the prior knowledge of the EM environment for constructing the radiative-reaction correction, their general applicability to a wide range of nanophotonic systems can be limited.

Here, we establish a direct Maxwell--Schr\"odinger scheme which enables first-principles or semi-empirical prediction of molecular spontaneous emission in a wide range of dielectric media. To avoid unstable radiative self-reaction when propagating Maxwell's equations, we employ molecular polarizations that suppress the spurious longitudinal component responsible for severe numerical fluctuations. Although spontaneous emission is usually regarded as a quantum process, this semiclassical scheme can accurately capture the radiative emission rate under the weak excitation limit \cite{Jaynes1963,Li2018Spontaneous,Chen2018Spontaneous}. With this approach, we demonstrate its broad applicability from atomic and molecular spontaneous emission in vacuum to enhanced radiative decay in both optical and plasmonic nanocavities.

\textit{Theoretical framework}---In this self-consistent Maxwell-Schr\"odinger scheme, the EM dynamics follow classical Maxwell's equations \cite{Griffiths1999}:
\begin{subequations}
\begin{align}
\partial_t \mathbf{D}(\mathbf{r}, t)
&= \nabla \times \mathbf{H}(\mathbf{r}, t)
- \mathbf{J}_{\mathrm{mol}}(\mathbf{r}, t),
\\[6pt]
\partial_t \mathbf{B}(\mathbf{r}, t)
&= - \nabla \times \mathbf{E}(\mathbf{r}, t).
\end{align}
\end{subequations}
Here, $\mathbf{E}$, $\mathbf{D}$, $\mathbf{H}$ and $\mathbf{B}$ denote the electric field, electric displacement, magnetic field and magnetic flux density, respectively, and $\mathbf{J}_{\rm mol}=\gamma \partial\mathbf{P}_{\rm mol}/\partial t$ is the molecular current density \cite{Sukharev2017, Li2018Spontaneous}, with $\gamma$ defined as a rescaling factor. The physical reality requires $\gamma=1$, and introducing  $\gamma>1$ accelerates the spontaneous emission dynamics, which will be discussed in more detail around Fig. \ref{fig:Figure2}. The propagation of $\mathbf{E}$, $\mathbf{D}$, $\mathbf{H}$ and $\mathbf{B}$ fields are efficiently solved using the finite-difference time-domain (FDTD) approach as implemented in MEEP  \cite{Yee1966, Oskooi2010}. The electronic dynamics are evolved using either RT-TDDFT implemented in CP2K \cite{Kühne2020} or more efficient density-functional tight-binding (RT-TDDFTB) implemented in DFTB+ \cite{Bonafe2020, Hourahine2020} [Supplementary Material (SM) Fig. S1]. In both cases, the electronic density $\hat{\mathbf{P}}^{\rm e}$ is propagated via
\begin{equation}\label{eq:electron_dynamics}
    \frac{d}{dt} \hat{\mathbf{P}}^{\rm e} = -\frac{\mathrm{i}}{\hbar} [\hat{\mathbf{F}}^{\rm e} + \hat{\mathbf{V}}^{\rm e},\  \hat{\mathbf{P}}^{\rm e}],
\end{equation}
represented under the orthogonal atomic orbital basis, with the electronic Kohn--Sham matrix $\hat{\mathbf{F}}^{\rm e}$ and light-matter coupling 
$\hat{\mathbf{V}}^{\rm e}
= - \gamma \int_{\Omega}
\mathrm{d}\mathbf{r}\,
\mathbf{E}_{\perp}(\mathbf{r}, t)
\cdot
\hat{\mathbf{P}}_{\mathrm{mol}}(\mathbf{r}) $.
At each time step, the FDTD solver provides the local electric field to the electronic dynamics, which in turn returns time derivative of mean-field molecule dipole ($\mathbf{P}_{\mathrm{mol}}(\mathbf{r}, t) = 
 \langle \hat{\mathbf{P}}_{\mathrm{mol}}(\mathbf{r}) \rangle
$
) to construct $\mathbf{J}_{\mathrm{mol}}$.
The EM and molecular subsystems communicate through the Internet socket interface within the \textsc{MaxwellLink} package \cite{Ji2026MaxwellLink}.

\begin{figure}[htbp]
\centering
\includegraphics[width=\columnwidth]{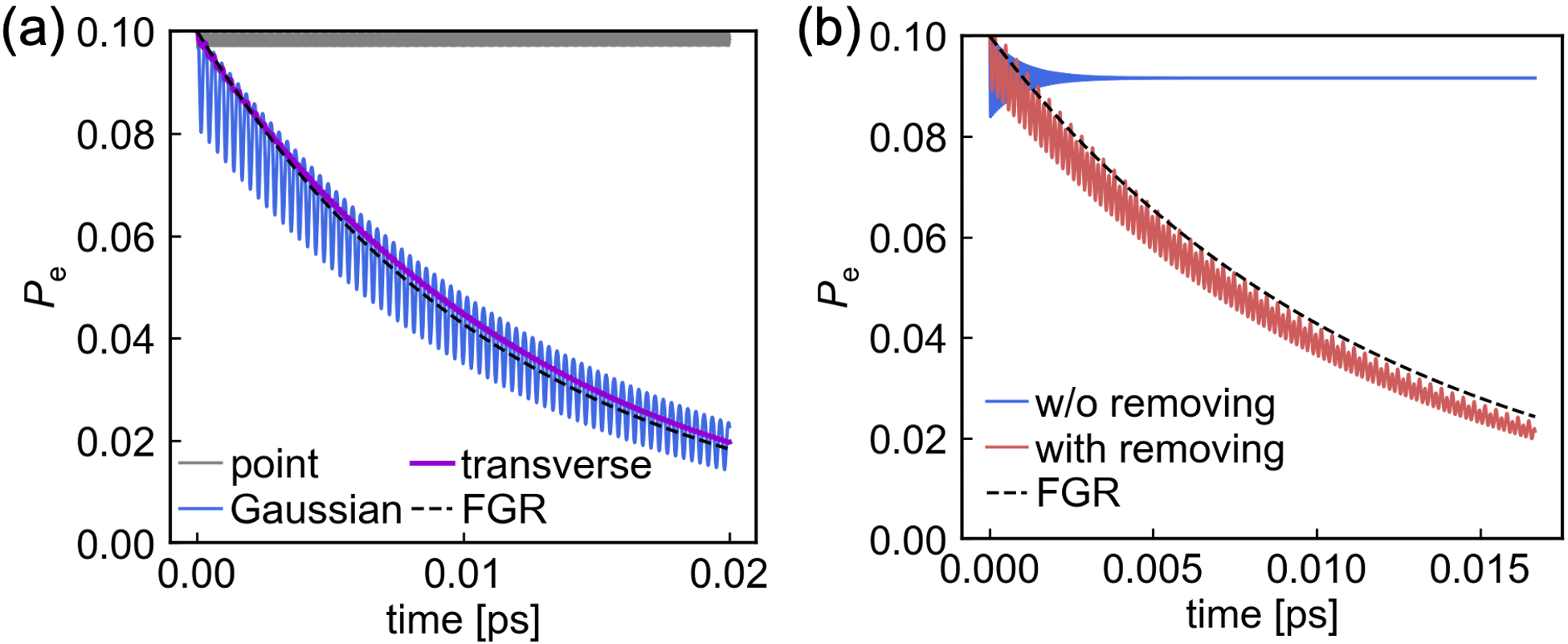}
\caption{Numerical stability of self-consistent Maxwell--Schr\"odinger simulations of spontaneous emission. (a) Excited-state population decay of a TLS in 3D vacuum using point-dipole, Gaussian  and transverse  polarizations, compared with the FGR prediction.  (b) Decay dynamics for a TLS spanning a one nanometer cell using a spatially localized Gaussian polarization with  and without  removing the static self-field.}
\label{fig:Figure1}
\end{figure}

\begin{figure}[htbp]
\centering
\includegraphics[width=\columnwidth]{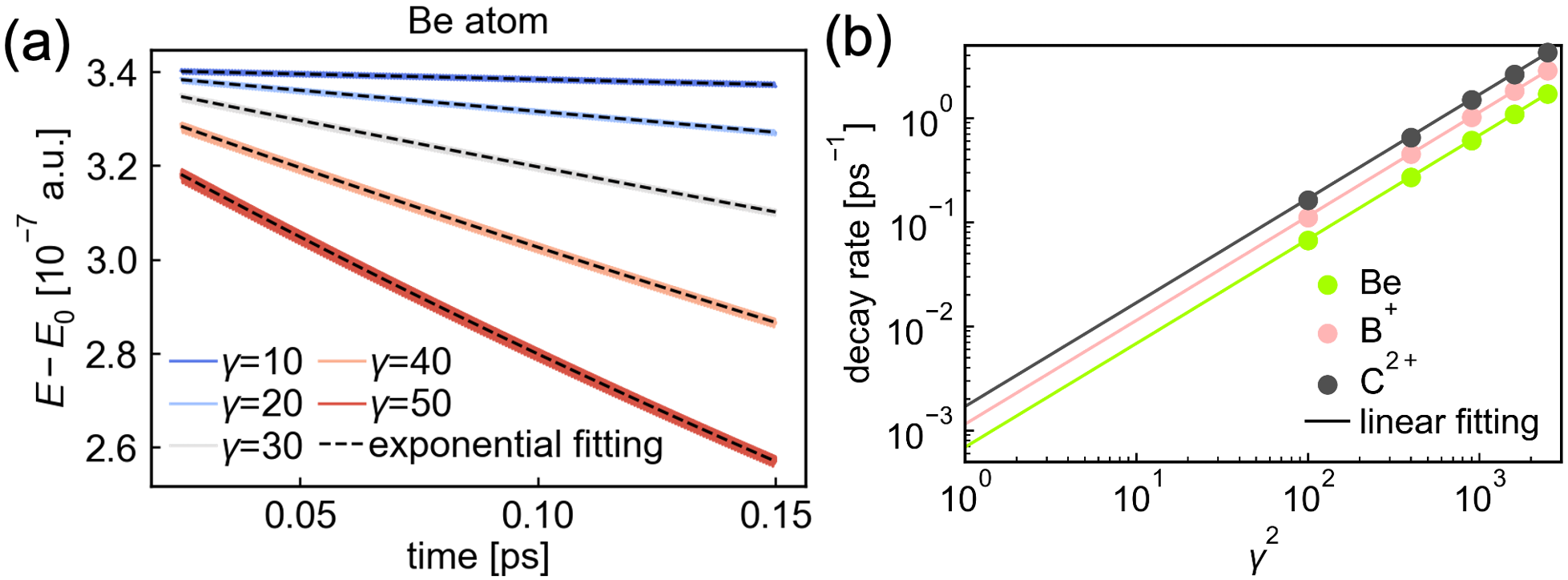}
\caption{Spontaneous decay of atoms in vacuum simulated by self-consistent FDTD--RT-TDDFT. (a) Electronic energy (relative to the ground state) evolution for the Be atom with different rescaling factors $\gamma$. (b) Decay rates, fitted from energy evolution, as a function of  $\gamma^2$ for Be, B$^+$ and C$^{2+}$. Solid lines show linear fits for three species, whose extrapolated values at $\gamma=1$ correspond to the physical decay rates.}
\label{fig:Figure2}
\end{figure}

\textit{Results}---To suppress the well-known numerical instabilities for evaluating spontaneous emission dynamics, we exclude the longitudinal component of $\hat{\mathbf{V}}^{\rm e}$ when numerically evaluating the light-matter integral by two means, the descriptions of which are provided in SM Section I.D as well as Figs. S2 and S3. In brief, for simulations in vacuum or optical cavities in which the molecular distribution is compatible with the emitter wavelength, we use a transverse polarization $\hat{\mathbf{P}}_{\mathrm{mol}}^{\perp}(\mathbf{r})$ constructed by applying the Helmholtz transverse projector in reciprocal space to the molecular polarization distribution $\hat{\mathbf{P}}_{\mathrm{mol}}(\mathbf{r})$.  As shown in Fig. \ref{fig:Figure1}a, using the transverse polarization (solid magenta) closely matches the Fermi's golden rule (FGR, dashed black) rate of spontaneous emission for an electronic two-level system (TLS) in three-dimensional (3D) vacuum, while the point-dipole approximation (solid gray) predicts utterly wrong radiative decay. The Gaussian polarization previously proposed by Li \textit{et al.} (solid blue) also suffers from large unphysical oscillations \cite{Li2018Spontaneous}.  

For molecules confined to nanometer length-scale regions, the spatially extended transverse polarization \cite{Cohen-Tannoudji1997} becomes incompatible with the limited simulation domain. We therefore take an alternative strategy by performing a reference FDTD simulation only once to determine the static field generated by the molecular polarization that causes strong numerical fluctuations, $\mathbf{E}_{\rm{self}}^{\rm{static}}(t)$, which is then subtracted from $\hat{\mathbf{V}}^{\rm e}$ during the subsequent self-consistent simulations. Fig. \ref{fig:Figure1}b provides the associated spontaneous emission dynamics for the TLS in vacuum using two-dimensional (2D) cylindrical coordinates. Clearly, this approach (solid red) yields spontaneous decay in good agreement with the FGR, while a spatially confined Gaussian polarization without this removal (solid blue) also fails drastically. 

To validate our self-consistent Maxwell--Schr\"odinger scheme beyond TLSs, following Ref. \citenum{Bustamante2021}, we study spontaneous emission rates for a set of atoms and ions. Fig. \ref{fig:Figure2}a presents coupled FDTD--RT-TDDFT calculations for a single Be atom in 3D vacuum after pulse excitation of the 2$s^2$ $\rightarrow$ 2$s$2$p(^1P)$ transition. The recorded atomic energy decays exponentially due to radiative emission. Importantly, to accelerate the decay, we use a sweep of enlarged rescaling factors $\gamma$ to enhance the light-matter coupling. As shown in Fig. \ref{fig:Figure2}b, the $\gamma\rightarrow 1$ limit of the  fitted decay lifetime, 1.47 ns, yields good agreement with the experimental lifetime (1.77--2.5 ns) \cite{WANG2018}.  The same procedure is further applied to the Be isoelectronic species B$^+$ and C$^{2+}$, producing radiative lifetimes of 0.88 and 0.60 ns, respectively. These results agree remarkably well with the available beam-foil measurements of $0.86\pm{0.07}$ and $0.57\pm{0.02}$ ns \cite{BASHKIN1985, Reistad1986}. Note that the linear relation for the fitted decay rates versus $\gamma^2$ suggests that all calculations are performed under the perturbative limit where the FGR (which depends quadratically on the coupling) remains valid.

\begin{figure}[htbp]
\centering
\includegraphics[width=0.9\columnwidth]{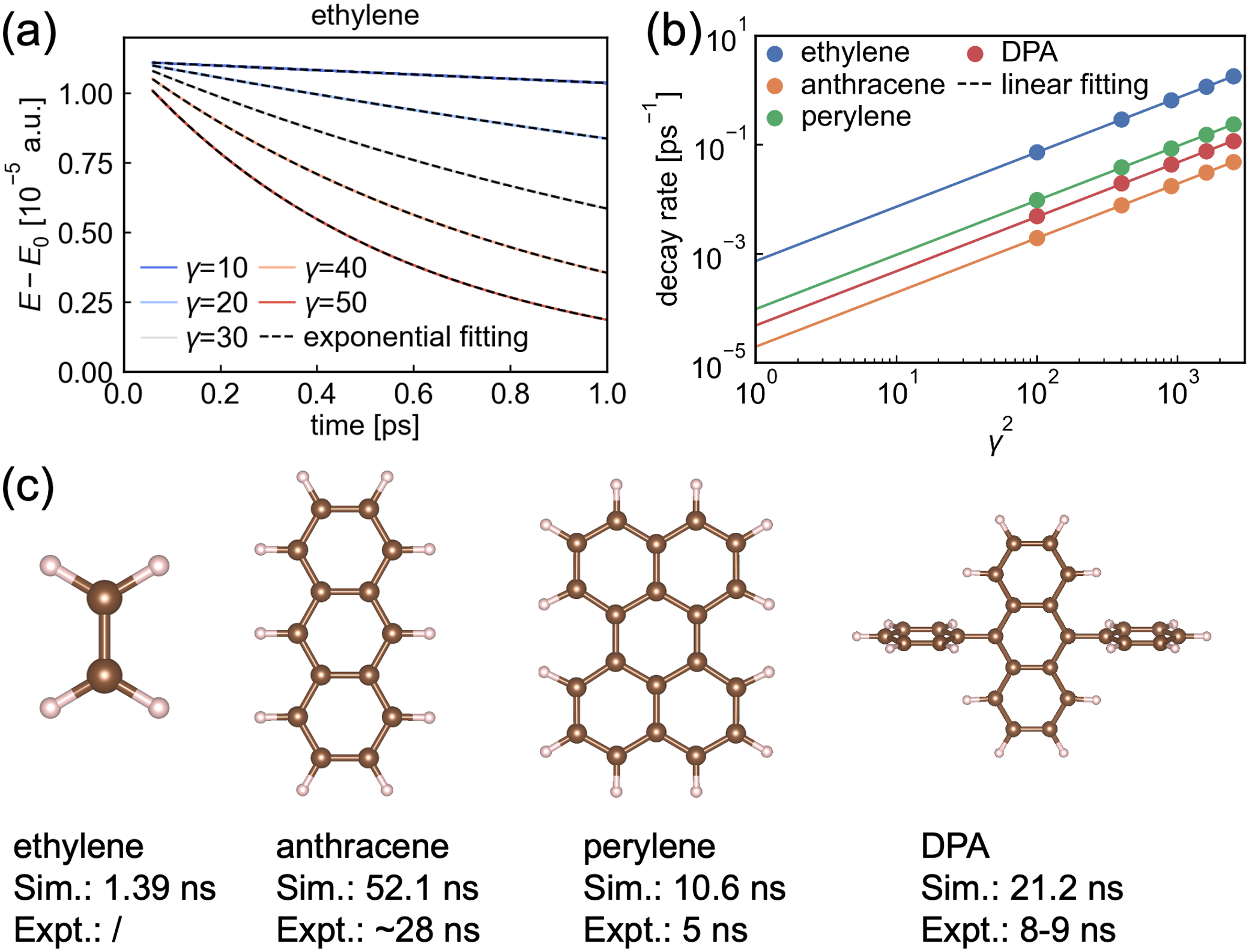}
\caption{Spontaneous decay of molecules in vacuum simulated by coupled FDTD--RT-TDDFTB. (a) Electronic energy  dynamics for the ethylene molecule with different $\gamma$. (b) Decay rates as a function of  $\gamma^2$ for ethylene, anthracene, perylene and DPA molecules. (c) Chemical structures of four molecules studied in this work and their radiative lifetimes from experiments \cite{morris1976fluorescence, Ware1965, Green2015} and simulations.}
\label{fig:Figure3}
\end{figure}

We further use this approach to compute the radiative lifetimes for a set of organic molecules.  Fig.~\ref{fig:Figure3}(a) shows the electronic energy evolution of a single ethylene molecule in 3D vacuum for different rescaling factors $\gamma$.
Here, for efficient calculations, we use the semi-empirical RT-TDDFTB for propagating 1 ps electronic dynamics.
The resulting radiative lifetime of ethylene is 1.39 ns [Figs.~\ref{fig:Figure3}(b) and (c)], in good agreement with the value of 1.13 ns obtained from corresponding FDTD--RT-TDDFT simulations (SM Fig. S4). This agreement supports the reliability of RT-TDDFTB for describing molecular spontaneous decay. We additionally simulate the radiative lifetimes of three organic molecules: anthracene, perylene and 9,10-diphenylanthracene (DPA). Compared with available experimental measurements \cite{morris1976fluorescence, Ware1965, Green2015}, simulated radiative lifetimes are roughly twice the experimental values. The discrepancy likely stems primarily from two sources. First, experimental lifetimes are measured in liquid solvents or solid polymer films, whereas our calculations are performed in vacuum. Second, the RT-TDDFTB approximation may introduce systematic errors in the description of molecular electronic structures.

\begin{figure}[htbp]
\centering
\includegraphics[width=\columnwidth]{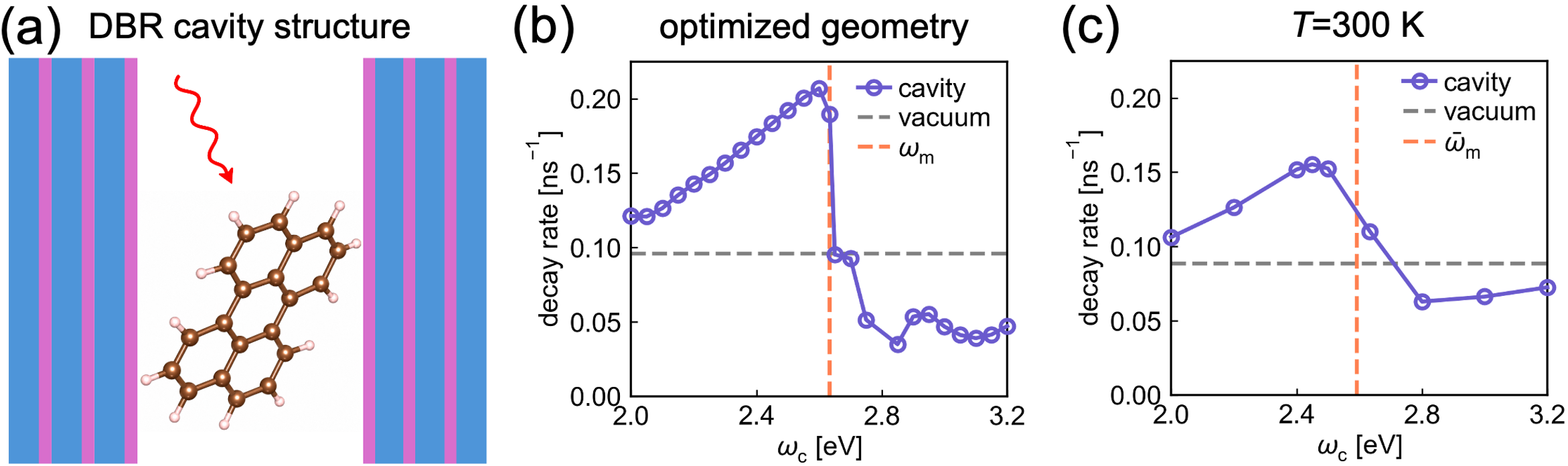}
\caption{Spontaneous decay of the perylene molecule in the DBR cavity. (a) Illustration of the simulation setup. (b) Decay rates of an optimized perylene molecule as a function of cavity frequency $\omega_{\rm c}$. The decay rate in vacuum (gray dashed line) and molecular $\text{S}_0\rightarrow \text{S}_1$ excitation $\omega_{\rm m}$ (vertical red line) are displayed for comparison. (c) Corresponding spontaneous decay rates as a function of $\omega_{\rm c}$ when each data point is averaged over 80 molecular geometries sampled from an NVT trajectory at 300 K.}
\label{fig:Figure4}
\end{figure}

The real advantage of our self-consistent scheme is the capability of conveniently simulating molecular radiative emission in complex dielectric environments. We next use the coupled FDTD--RT-TDDFTB approach to simulate a single perylene molecule in an optical cavity formed by a pair of distributed Bragg reflectors (DBRs) in 3D, as depicted in Fig.~\ref{fig:Figure4}(a). Detailed cavity setting parameters are shown in Fig. S5 of the SM. Fig.~\ref{fig:Figure4}(b) shows the spontaneous decay rates of a single perylene molecule with optimized geometry as a function of the cavity frequency $\omega_{\mathrm{c}}$. The decay rate reaches its maximum when $\omega_{\mathrm{c}}$ is resonant with molecular lowest $\text{S}_0\rightarrow \text{S}_1$ excitation $\omega_{\rm m}$. The simulated Purcell factor \cite{purcell1995spontaneous}  is 2.2, comparable to the reported value of $2.4\pm{0.2}$ for blue phosphorescent OLEDs \cite{zhao2024stable}.

To further account for thermal structural fluctuations under experimentally relevant conditions, we sample 80 molecular geometries from an NVT trajectory at 300 K. The excitation energy and vacuum decay rate distributions of these geometries are shown in Fig. S6 of the SM. Cavity-modified decay rates for these geometries are simulated and averaged for different cavity frequencies [Fig.~\ref{fig:Figure4}(c)], where the maximum Purcell factor reduces to 1.8 due to this thermal average. As a moderate enhancement can lead to several-fold improvements in OLED operational lifetimes \cite{zhao2024stable}, a predictive theoretical framework for molecular radiative lifetimes in optical cavities becomes essential for improving device performances.

One interesting observation in molecular Purcell effects is that, at $T$=300 K, the maximum cavity Purcell factor occurs at a red-detuned $\omega_{\rm c}$ with respect to the apparent molecular excitation transition $\bar{\omega}_{\rm m}$ averaged over thermally fluctuating molecular geometries [$\omega_{\rm c}  - \bar{\omega}_{\rm m}\sim -0.14$ eV, Fig.~\ref{fig:Figure4}(c)]. This result suggests that the optimal cavity enhancement may require a red detuning at room temperature instead of exact resonance with molecular excitation. This shift can be understood from the asymmetrical dependence of decay rates on the cavity frequency. For a fixed molecular geometry, as shown in Fig.~\ref{fig:Figure4}(b), when $\omega_{\mathrm{c}}$ is tuned below $\omega_{\rm m}$, the decay rate decreases gradually towards its vacuum value. In contrast, when $\omega_{\mathrm{c}}$ exceeds $\omega_{\rm m}$, the decay rate sharply drops to half of the vacuum value. As a result, for an ensemble of thermally distributed molecular geometries, this asymmetry requires a red-detuned $\omega_{\mathrm{c}}$ with respect to molecular average transition frequency $\bar{\omega}_{\rm m}$ to yield the largest apparent cavity enhancement. The corresponding decay rate distributions for different $\omega_{\mathrm{c}}$ are shown in Fig. S7 of the SM. This result suggests that molecules differ substantially from atoms for Purcell effects due to their flexible geometries at room temperature.

\begin{figure}[htbp]
\centering
\includegraphics[width=0.9\columnwidth]{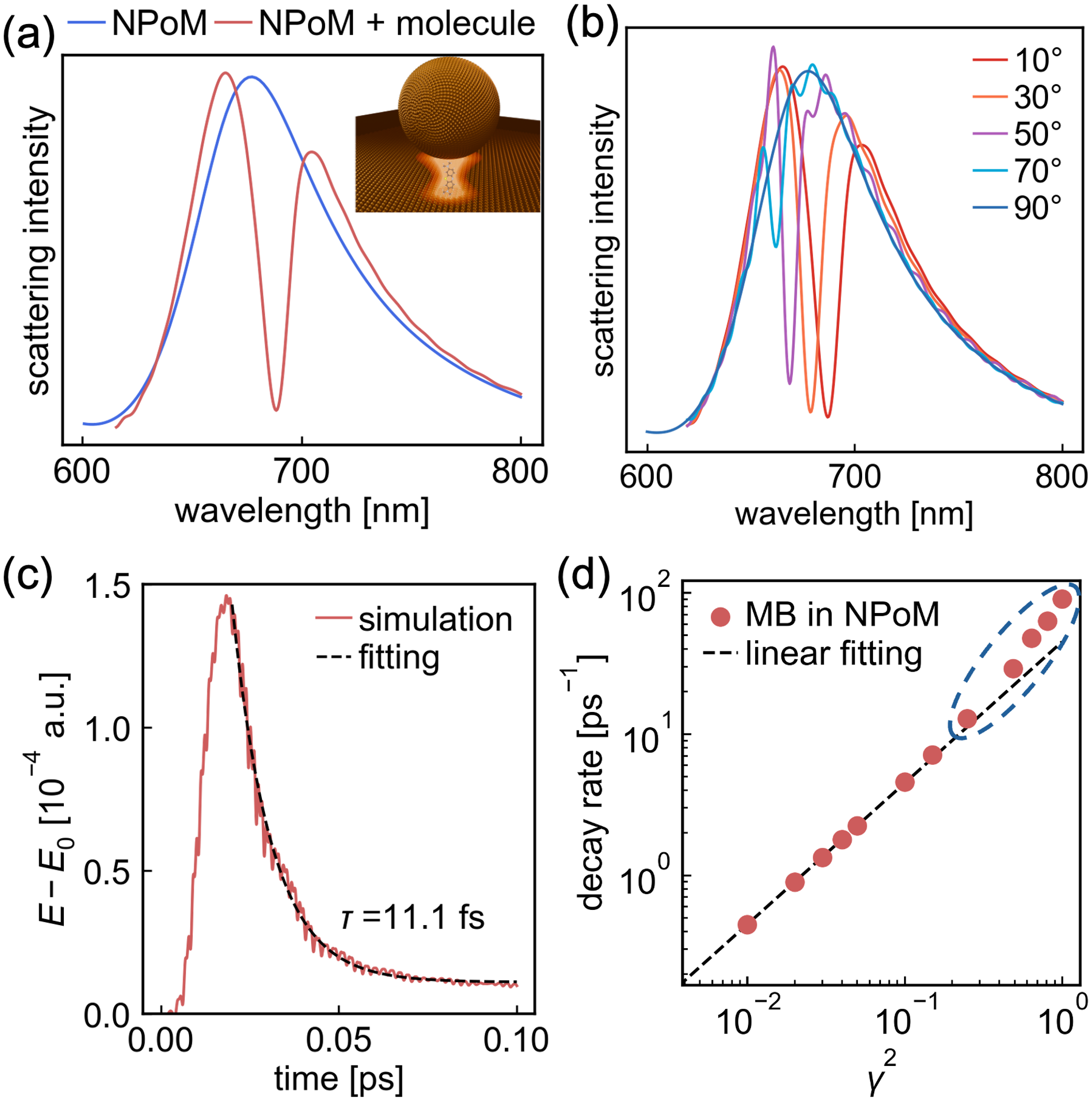}
\caption{Strong coupling of a single methylene blue molecule in a plasmonic NPoM nanocavity. (a) Scattering spectra of an isolated NPoM nanocavity (blue) and a methylene blue molecule perpendicularly placed in the NPoM nanocavity (red). (b) Scattering spectra of a methylene blue molecule placed in the NPoM nanocavity with different orientations. (c) Energy evolution dynamics of a methylene blue molecule excited by a Gaussian pulse in the NPoM nanocavity, where the trajectory after turning off the external field is used for exponential fitting to determine the decay rate. (d) Decay rates as a function of  $\gamma^2$ for a methylene blue molecule in the NPoM cavity.}
\label{fig:Figure5}
\end{figure}

Finally, we extend our non-perturbative Maxwell--Schr\"odinger framework to model single-molecule plasmonic strong coupling experiments  in the NPoM nanocavity [inset panel of Fig.~\ref{fig:Figure5}(a)] \cite{Chikkaraddy2016, baumberg2019extreme}.
A methylene blue molecule is placed at the center of the gap to maximize its coupling to the cavity mode; see SM Fig. S8 for the nanocavity setting. 

This experiment provides a stress test of our approach. On the one hand, single-molecule strong coupling requires the description of spontaneous emission beyond perturbation theory. On the other hand, the strongly inhomogeneous EM fields created by this NPoM cavity also demand high spatial resolution in FDTD for considering both the molecular length scale (sub-nanometer) and the cavity length scale (> 600 nanometers). Such a large simulation domain combined with the high spatial resolution would make the 3D FDTD simulation prohibitively expensive. To this end, we further exploit the cylindrical symmetry of the NPoM geometry and perform FDTD simulations in 2D cylindrical coordinates, thus significantly reducing the computational cost. Convergence tests of numerical parameters are shown in Figs. S9 and S10 of the SM.

Fig.~\ref{fig:Figure5}(a) compares the scattering spectra of bare NPoM nanocavity and the nanocavity containing a single methylene blue molecule. The cavity resonance corresponding to the 40 nm-diameter nanoparticle is located at $\omega_{\mathrm{p}}$ = 677 nm (1.83 eV), in good agreement with experimental measurements of $660\pm{10}$ nm \cite{Chikkaraddy2016}. When this molecule is perpendicularly placed at the center of the NPoM nanogap, the spectrum splits into two peaks, corresponding to the formation of plasmon-exciton polariton states \cite{paoletta2024plasmon}. The spectral Rabi splitting is 104 meV, consistent with experimentally estimated range of $80 \sim 95$ meV \cite{Chikkaraddy2016}. We further rotate the molecular orientation with respect to the cavity. As shown in Fig.~\ref{fig:Figure5}(b), the Rabi splitting decreases as the rotation angle increases and finally vanishes when the molecule is parallel to the mirror, also consistent with experimental spectra with different molecular orientations \cite{Chikkaraddy2016}.

We then examine the energy relaxation of the methylene blue molecule perpendicularly placed in the nanocavity. Fig.~\ref{fig:Figure5}(c) shows that molecular electronic energy decays on an ultrafast time scale, yielding a radiative emission lifetime of 11.1 fs from an exponential fit. Remarkably, this lifetime is approximately $3.4\times10^6$ times smaller than the vacuum result (38.2 ns). This lifetime acceleration is due mainly to coherent Rabi oscillations coupled to strong plasmonic absorption. This timescale is comparable to that of nonadiabatic processes, indicating that the radiative emission (or electrodynamical relaxation) process can directly compete with molecular vibronic coupling pathway in plasmonic nanocavities. Such competition is relevant to plasmon-assisted photochemistry and photocatalysis \cite{halas2011plasmons, mukherjee2014hot, Linic2015, zhang2018surface}, where the branching between these two pathways can influence molecular dynamics and chemical reactions.

Fig.~\ref{fig:Figure5}(d) further demonstrates the decay rate as a function of the rescaling factor $\gamma$. Here, instead of increasing $\gamma$ as above, we reduce $\gamma$ to examine how the spontaneous emission may vary for an artificial molecule with a reduced transition dipole moment.  For $\gamma^2 \le 10^{-1}$, the decay rate follows a perfect linear dependence on $\gamma^2$, indicating the perturbative regime. Interestingly, with further increased $\gamma^2$, the decay rate gradually deviates from the linear relation (blue elliptical area), suggesting that the system starts to enter the strong coupling regime and requires a non-perturbative treatment of molecular electrodynamical relaxation \cite{kuisma2022ultrastrong}.

\textit{Conclusion}---Our self-consistent Maxwell--Schr\"odinger framework resolves a longstanding computational bottleneck prohibiting accurate dynamical simulations of molecular spontaneous emission in complex EM environments. The versatility of this framework is evidenced by its successful applications to weak and strong light-matter coupling regimes.
As a future extension, this framework can naturally incorporate nuclear dynamics to study photophysics and photochemistry in a wide range of dielectric environments.

\begin{acknowledgments}
\textit{Acknowledgments}---This material is based upon work supported by the U.S. National Science Foundation under Grant No. CHE-2502758. This work used the Anvil HPC at Purdue University through allocation CHE250091 from the Advanced Cyberinfrastructure Coordination Ecosystem: Services \& Support (ACCESS) program \cite{Boerner2023}, which is supported by U.S. National Science Foundation grants \#2138259, \#2138286, \#2138307, \#2137603, and \#2138296.
\end{acknowledgments}

\end{document}


\begin{center}
{\Large \bfseries Supplemental Material}\\[4pt]
{\Large for}\\[6pt]
{\Large\bfseries \sititle}\\[10pt]

{\large \siauthors}\\[4pt]
{\siaffiliation}
\end{center}

\vspace{1em}

\tableofcontents

\thispagestyle{empty}

\vfill
\noindent $^{*}$\href{mailto:\siemail}{\siemail}

\clearpage

\section{Theory}
\label{sec:theory}
\subsection{Self-consistent FDTD--RT-TDDFT(B) framework}
\label{sec:FDTD-TDDFTB}

The electromagnetic and electronic dynamics are propagated self-consistently within the \textsc{MaxwellLink} framework \cite{Ji2026MaxwellLink}. The electromagnetic field follows the classical Maxwell's equations \cite{Griffiths1999}:
\begin{subequations}
\begin{align}
\partial_t \mathbf{D}(\mathbf{r},t)
&=\nabla \times \mathbf{H}(\mathbf{r},t)-
\mathbf{J}_{\mathrm{mol}}(\mathbf{r},t),
\\
\partial_t \mathbf{B}(\mathbf{r},t)
&=-\nabla \times \mathbf{E}(\mathbf{r},t).
\end{align}
\end{subequations}
Here, $\mathbf{E}$, $\mathbf{D}$, $\mathbf{H}$ and $\mathbf{B}$ denote electric field, electric displacement, magnetic field and magnetic flux density, respectively. The displacement field $\mathbf{D}$ is defined as:
\begin{equation}
\mathbf{D}(\mathbf{r}, \omega)
=\epsilon_0\epsilon_r^{*}(\mathbf{r}, \omega)\mathbf{E}(\mathbf{r}, \omega),
\end{equation}
where $\epsilon_0$ is the vacuum permittivity, and $\epsilon_r^{*}(\mathbf{r}, \omega)$ denotes the relative permittivity of the dielectric media. Similarly, $\mathbf{B}$ field is related to $\mathbf{H}$ field via $\mathbf{B}$=$\mu$$\mathbf{H}$, where $\mu$ is the magnetic permeability. $\mathbf{E}$, $\mathbf{D}$, $\mathbf{H}$ and $\mathbf{B}$ fields can be efficiently propagated using the standard finite-difference time-domain (FDTD) algorithm \cite{Yee1966}.

The molecular response enters Maxwell's equations through the current density \cite{Sukharev2017, Li2018Spontaneous}:
\begin{equation}
\mathbf{J}_{\mathrm{mol}}(\mathbf{r}, t)=
\frac{\partial \mathbf{P}_{\mathrm{mol}}(\mathbf{r}, t)}{\partial t},
\end{equation}
where $\mathbf{P}_{\mathrm{mol}}$ is the molecular polarization. In the present framework, the molecular polarization is expressed as:
\begin{equation}
\mathbf{P}_{\mathrm{mol}}(\mathbf{r}, t)=
\sum_{i=x,y,z}
\gamma \mu_{i}(t)\kappa_{i}(\mathbf{r})\mathbf{e}_{i},
\label{eq:molecular-polarization}
\end{equation}
where $\mu_i(t)$ is the time-dependent molecular dipole moment along direction $i=x, y, z$, $\kappa_{i}(\mathbf{r})$ describes its spatial distribution on the FDTD grid, and $\gamma$ is the rescaling factor for the light-matter interaction ($\gamma=1$ corresponds to the physical coupling strength). The specific polarization distribution functions employed in this work will be discussed in Section~\ref{sec:polarization-function}.

From Eq.~\eqref{eq:molecular-polarization}, the interaction between the molecular subsystem and the electromagnetic field is defined as:
\begin{equation}
V_{\mathrm{coupling}}=
-\int_{\Omega_\mathrm{m}} \mathrm{d}\mathbf{r}\,
\mathbf{E}_{\perp}(\mathbf{r},t)\cdot\mathbf{P}_{\mathrm{mol}}(\mathbf{r},t) = -\int_{\Omega_\mathrm{m}} \mathrm{d}\mathbf{r}\,
\mathbf{E}(\mathbf{r},t)\cdot\mathbf{P}^{\perp}_{\mathrm{mol}}(\mathbf{r},t),
\label{eq:light-matter-coupling-raw}
\end{equation}
where $\mathbf{E}_{\perp}(\mathbf{r},t)$ denotes the transverse component of the classical electric field $\mathbf{E}(\mathbf{r},t)$. Because it is time-consuming to evaluate $\mathbf{E}_{\perp}(\mathbf{r},t)$ from the FDTD propagation of $\mathbf{E}(\mathbf{r},t)$ at each time step (which involves a spatial Fourier transform), we can use a transverse polarization density $\mathbf{P}^{\perp}_{\mathrm{mol}}(\mathbf{r},t)$ instead for evaluating the coupling. As a result,
\begin{equation}
V_{\mathrm{coupling}}=
-\int_{\Omega_\mathrm{m}} \mathrm{d}\mathbf{r}\,
\mathbf{E}(\mathbf{r},t)\cdot\mathbf{P}^{\perp}_{\mathrm{mol}}(\mathbf{r},t)
=-\gamma\sum_{i=x,y,z}\mu_i(t) E_i^{\mathrm{eff}}(t),
\label{eq:light-matter-coupling}
\end{equation}
where the effective local electric field is:
\begin{equation}
E_i^{\mathrm{eff}}(t)=
\int_{\Omega_\mathrm{m}} \mathrm{d}\mathbf{r}\,
E_i(\mathbf{r},t)\kappa_{i}^{\perp}(\mathbf{r}).
\label{eq:effective-e-field}
\end{equation}
The $\mu_i(t) E_i^{\mathrm{eff}}(t)$ term resembles the point-dipole approximation, but the above derivation indicates that a raw FDTD E-field at the molecular location is not sufficient. Instead, an effective E-field $E_i^{\mathrm{eff}}(t)$ regularized by Eq. \eqref{eq:effective-e-field} becomes necessary for properly dealing with molecular radiative emission.

\begin{figure}[htbp]
\centering
\includegraphics[width=0.8\columnwidth]{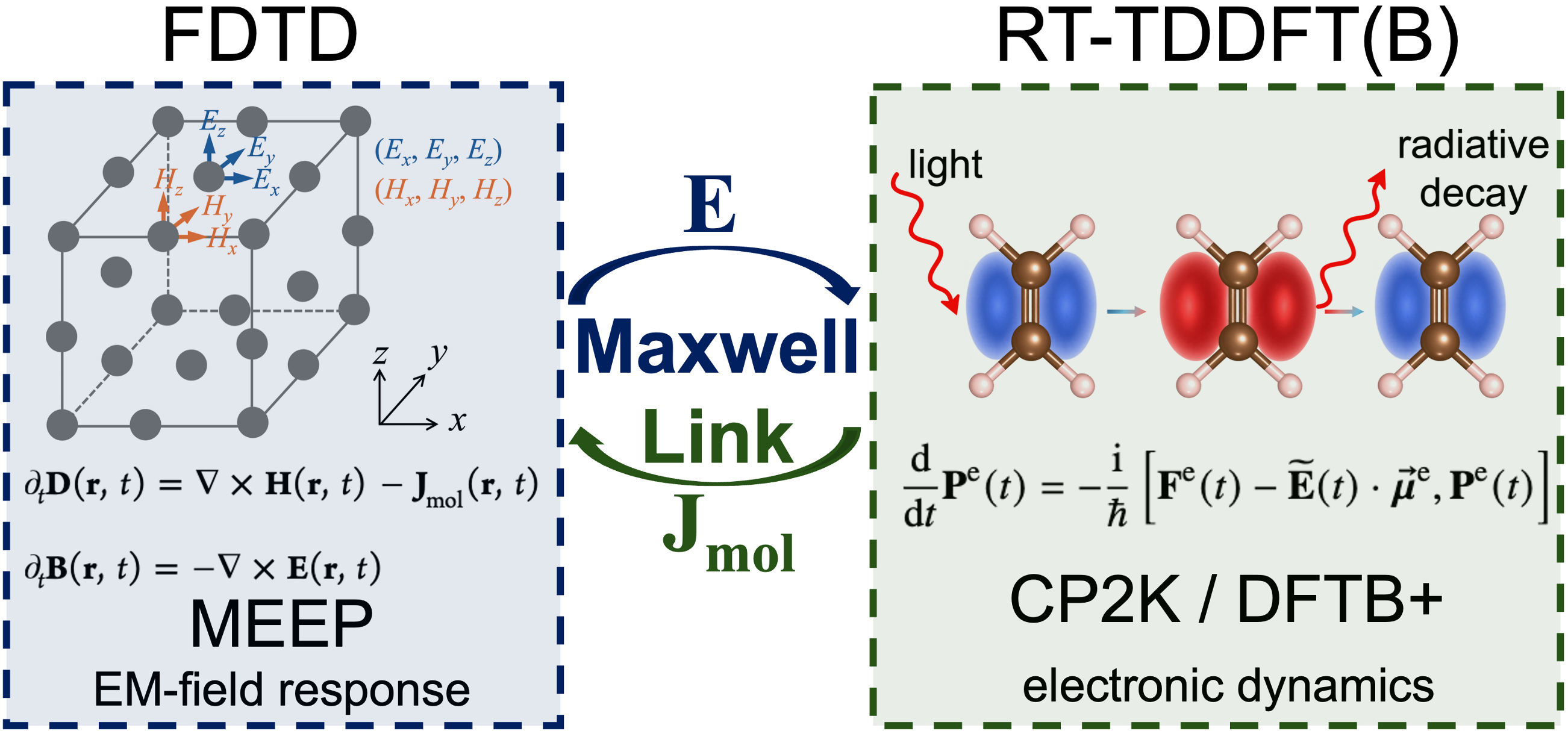}
\caption{Schematic diagram of self-consistent FDTD--RT-TDDFT(B) simulations of molecular spontaneous decay in dielectric media.}
\label{fig:Figure1}
\end{figure}

As shown in Fig. \ref{fig:Figure1}, at each propagation step, the FDTD solver calculates $\mathbf{E}^{\mathrm{eff}}(t)$=[$E_x^{\mathrm{eff}}(t)$, $E_y^{\mathrm{eff}}(t)$, $E_z^{\mathrm{eff}}(t)$] and transfers it to the electronic dynamics driver through the TCP/UNIX socket interface in \textsc{MaxwellLink}. The electronic driver then evolves electronic dynamics under the field interaction and returns the time derivative of the molecular dipole $\dot{\boldsymbol{\mu}}_m=[\dot{\mu}_m^x, \dot{\mu}_m^y, \dot{\mu}_m^z]$, which is used to construct $\mathbf{J}_\mathrm{mol}$ for the subsequent FDTD step. Specifically, the time-dependent molecular dipole moment generates an electromagnetic response, which subsequently acts back on the electronic dynamics and enables radiative energy dissipation using semiclassical dynamics.

\subsection{RT-TDDFT}
\label{sec:RT-TDDFT}
Within real-time time-dependent density functional theory (RT-TDDFT), the electronic dynamics are propagated through the time-dependent Kohn-Sham equations \cite{Runge1984, hanasaki2025development}:
\begin{equation}
\mathrm{i}\frac{\partial}{\partial t}\varphi_i(\mathbf{r},t)
=\left(\hat{T}+v_{\mathrm{ext}}(\mathbf{r},t)+v_{\mathrm{H}}(\mathbf{r},t)+v_{\mathrm{xc}}(\mathbf{r},t)\right)
\varphi_i(\mathbf{r},t),
\end{equation}
where $\hat{T}$ is the kinetic energy operator, and $v_{\mathrm{ext}}(\mathbf{r},t)$, $v_{\mathrm{H}}(\mathbf{r},t)$ and $v_{\mathrm{xc}}(\mathbf{r},t)$ are the external, Hartree, and exchange-correlation potentials, respectively. $v_{\mathrm{ext}}(\mathbf{r},t)$ includes the electron-nuclear interaction and the time-dependent light-matter coupling $-\mathbf{E}^{\mathrm{eff}}(t)\cdot\hat{\boldsymbol{\mu}}$, where $\mathbf{E}^{\mathrm{eff}}$ is the effective electric field obtained from the FDTD simulation and $\hat{\boldsymbol{\mu}}$ is the molecular dipole operator. The resulting time derivative of molecular dipole is then returned back for the subsequent FDTD step.

\subsection{RT-TDDFTB}
\label{sec:RT-TDDFTB}
Real-time time-dependent density functional tight-binding (RT-TDDFTB) provides an efficient approximation to RT-TDDFT by propagating the electronic density matrix within the density functional tight-binding framework \cite{Bonafe2020}. In second-order DFTB (DFTB2) \cite{Elstner1998}, the Kohn-Sham DFT energy is expanded around a reference density $n_0$ to the second order of density fluctuation. Within the self-consistent-charge DFTB (SCC-DFTB) approximation, the Hamiltonian matrix in the nonorthogonal atomic-orbital basis is written as:
\begin{equation}
H_{\mu\nu}=H_{\mu\nu}^{0}+\frac{1}{2}S_{\mu\nu}\sum_{C}\left(\gamma_{AC}+\gamma_{BC}
\right)\Delta q_C,\qquad\mu \in A,\ \nu \in B.
\end{equation}
Here, $H_{\mu\nu}^{0}$ is the reference Hamiltonian matrix element, $S_{\mu\nu}$ is the overlap matrix element, $\Delta q_C$ is the Mulliken charge variance on atom $C$, and $\gamma_{AC}$ describes the effective Coulomb interaction between charge fluctuations on atoms $A$ and $C$. Mulliken charges are obtained from the electronic density matrix $\rho$. 

The DFTB3 extension \cite{gaus2011dftb3} further includes the third-order density-fluctuation contribution through charge-dependent Hubbard parameters. Within the nonorthogonal basis, the real-time electronic dynamics obey the quantum von Neumann equation:
\begin{equation}
\dot{\rho}=-\mathrm{i}\left(S^{-1} H \rho-\rho H S^{-1}\right).
\end{equation}
The time derivative of molecular dipole is also evaluated from the propagated electronic density and is subsequently coupled to the FDTD simulation through the \textsc{MaxwellLink} framework described above.

\subsection{Polarization functions and the treatment of molecular self-field}
\label{sec:polarization-function}
The spatial polarization function $\kappa_{i}(\mathbf{r})$ in Eq.~\ref{eq:molecular-polarization} determines how the molecular dipole is represented on the FDTD grid and how the electric field is sampled in the light-matter interaction of Eqs.~\ref{eq:light-matter-coupling} and \ref{eq:effective-e-field}. The molecular polarization generates both transverse radiative fields and longitudinal self-fields. The former provides physical electromagnetic back-action responsible for radiative relaxation, whereas the latter can introduce large nonphysical numerical oscillations \cite{Li2018Spontaneous, Ji2026MaxwellLink}. Therefore, the treatment of the spatial polarization function is particularly significant for self-consistent simulations of spontaneous emission.

For evaluating Eq. \eqref{eq:effective-e-field}, the simplest approach is to directly apply the point-dipole polarization:
\begin{equation}
\kappa_i^{\mathrm{point}}(\mathbf{r})=
\delta(r_i-r_i^\mathrm{m})
\end{equation}
without any further treatment,
where $r_i^\mathrm{m}$ denotes molecular position along $i=x, y, z$ directions. Such a highly localized source can result in substantial numerical fluctuations in self-consistent simulations \cite{Ji2026MaxwellLink}. In our practical implementation in \textsc{MaxwellLink}, the local field is averaged over one FDTD grid cell when evaluating the light-matter coupling, which reduces but does not eliminate this numerical instability.

A smoother representation is the Gaussian polarization \cite{Li2018Spontaneous}. Along direction $i$, the analytical Gaussian kernel employed in this work can be expressed as:
\begin{equation}
\kappa_i^{\mathrm{G}}(\mathbf{r})=
\frac{(r_i-r_i^\mathrm{m})^2}{(2\pi)^{3/2}\sigma^5}\exp\left[
-\frac{|\mathbf{r}-\mathbf{r}_m|^2}{2\sigma^2}\right],
\end{equation}
where $\sigma$ controls the spatial width of the polarization distribution. This distribution substantially reduces the numerical instability of the point-dipole polarization. Nevertheless, it still contains both transverse and longitudinal components.

For simulations with sufficiently large cells, we further construct the transverse polarization from the Gaussian distribution. According to the Helmholtz decomposition, the polarization can be separated into transverse and longitudinal components in reciprocal space \cite{Cohen-Tannoudji1997}. The transverse component is obtained by applying the projection operator:
\begin{equation}
\widetilde{\kappa}_i^{\perp}(\mathbf{k})=
\left(\mathbf{I}-\frac{\mathbf{k}\mathbf{k}^{\mathrm{T}}}{|\mathbf{k}|^2}\right)\widetilde{\kappa}_i^{\mathrm{G}}(\mathbf{k}),
\end{equation}
followed by an inverse Fourier transform to real space. After this procedure, the longitudinal component associated with molecular polarization is mostly eliminated, while the transverse component responsible for radiative back-action is retained. In the implementation, this transverse kernel is numerically constructed using a fast Fourier transform only once at the beginning of the simulation, so the overall computational cost remains the same level as those using point-dipole or Gaussian polarization densities. Of course, a sufficiently large real-space region (with an extent comparable to a wavelength corresponding to the emitter frequency) is required to obtain a converged delocalized transverse distribution.

\begin{figure}[H]
\centering
\includegraphics[width=0.7\textwidth]{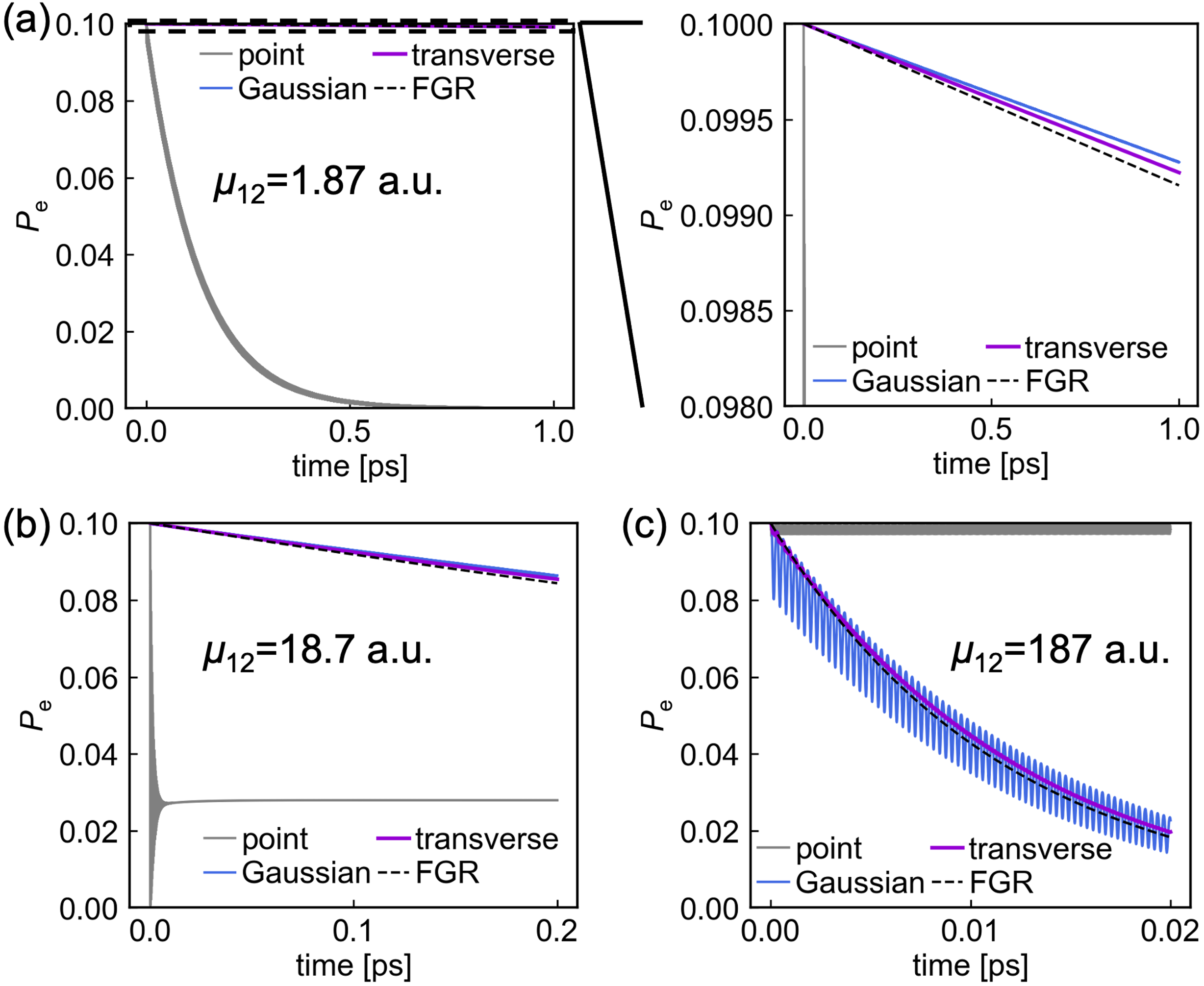}
\caption{Comparison of excited-state population decay of a TLS in vacuum using point, Gaussian and transverse polarizations, along with the FGR decay. The excitation energy of the TLS is fixed to 0.484 a.u., while the transition dipole moment $\mu_{12}$ is set to (a) 1.87, (b) 18.7 and (c) 187 a.u., respectively.}
\label{fig:polarization-test}
\end{figure}

The population evolution of a two-level system (TLS) in 3D vacuum using different polarization kernels is shown in Fig.~\ref{fig:polarization-test}. To accelerate spontaneous decay within a computationally accessible propagation time, we employed a large rescaling factor $\gamma$ in production simulations of the main text, which is equivalent to increasing the effective transition dipole moment. Consequently, Fig.~\ref{fig:polarization-test} examines much larger transition dipoles than realistic molecules to test the numerical stability of different polarization schemes. 

As shown in Fig.~\ref{fig:polarization-test},  the point-dipole polarization largely deviates from the 3D Fermi's golden rule (FGR) decay, arising from its strongly localized field. Gaussian polarization greatly improves the decay dynamics, although residual oscillations become pronounced with increased light-matter coupling. In contrast, the transverse polarization largely suppresses these oscillations and closely matches the FGR decay over the investigated coupling range.  These results demonstrate that removing the longitudinal component is important when a large rescaling factor is used to accelerate spontaneous emission dynamics. Here, the large dipole limit (Fig.~\ref{fig:polarization-test}c) is particularly relevant, which aims to describe spontaneous decay within 20 fs, which is usually the simulation time scale of RT-TDDFT. Because the transverse polarization captures spontaneous emission with this time scale, it becomes possible to interpolate the physical spontaneous decay rate following the recipe of Fig. 2 of the main text, i.e., through dividing the fitted decay rate by a factor of $\gamma^2$.

However, the spatially extended transverse kernel distribution is not suitable for strongly confined electromagnetic environments. This is particularly relevant to the NPoM nanocavity used in the main text, where a single molecule is placed in the one nanometer gap. In such cases, we use a spatially localized Gaussian polarization together with subtracting its static self-field. In detail, a cheap reference FDTD simulation is first performed to determine the static field generated by a molecular polarization with a unit dipole in a homogeneous medium (corresponding to the background dielectric medium of the molecule). This static field is stored as $E_{\mathrm{self},i}^{\mathrm{static}}$ with $i=x,y,z$ throughout the simulation. Then, as the real molecule has an associated mean-field dipole moment $\mu_i(t)$, the time-dependent static field vector can be computed on the fly by $\mathbf{E}_{\mathrm{self}}^{\mathrm{static}}(t) = \sum_i E_{\mathrm{self},i}^{\mathrm{static}} \mu_i(t) \mathbf{e}_i$.  During the subsequent self-consistent propagation, this contribution is subtracted from the effective field entering the light-matter interaction:
\begin{equation}
\mathbf{E}_{\mathrm{eff}}^{\mathrm{corr}}(t)
=\mathbf{E}_{\mathrm{eff}}(t)-
\mathbf{E}_{\mathrm{self}}^{\mathrm{static}}(t).
\end{equation}
This removal effectively eliminates the zero-frequency real component of self dyadic Green's function, which is the dominant contribution causing numerical instabilities.

\begin{figure}[H]
\centering
\includegraphics[width=1.0\textwidth]{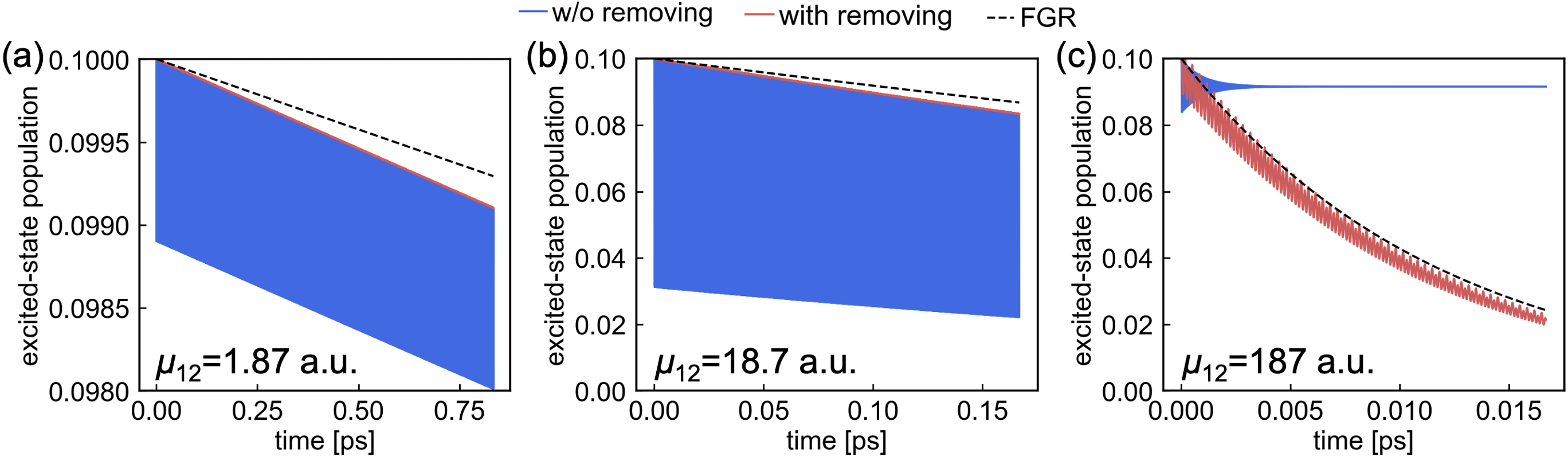}
\caption{Comparison of excited population decay of a TLS in vacuum using a spatially localized Gaussian polarization with and without removing static self-field, along with decay calculated from the FGR. The excitation energy of the TLS is fixed to 0.484 a.u., while the transition dipole moment $\mu_{12}$ is set to (a) 1.87, (b) 18.7 and (c) 187 a.u., respectively.}
\label{fig:remove-static-field-test}
\end{figure}

Fig.~\ref{fig:remove-static-field-test} benchmarks this alternative treatment using the same TLS model when the spatial extent of the molecule is confined in a one-nanometer-length box in the FDTD grid. Without subtracting the static self-field, the spatially localized Gaussian polarization exhibits increasing oscillations with enhanced transition dipole moments, and the resulting decay dynamics  largely deviate from FGR's prediction, especially under the limit of large transition dipole moments. After removing the static self-field, these oscillations are strongly suppressed and the population decay closely matches the FGR result, even for the largest transition dipole moment considered here.

\section{Simulation Details}
\label{sec:simulation-details}

The electromagnetic dynamics were propagated by FDTD as implemented in MEEP \cite{Oskooi2010}. Except for the nanoparticle-on-mirror (NPoM) nanocavity simulations discussed below, all FDTD simulations were performed in three-dimensional (3D) Cartesian coordinates. The MEEP time unit was set to 0.1 fs, corresponding to a length unit of approximately 30 nm. FDTD spatial resolution was set to 10, giving a grid spacing of approximately 3 nm. For atoms and molecules in vacuum, a cubic simulation cell of $8\times8\times8$ MEEP units ($240\times240\times240$ $\mathrm{nm}^3$) was used. For absorbing boundary conditions, a 3.0 MEEP units-thick perfectly matched layer (PML) was applied at both ends of the simulation cell to prevent unphysical reflections. For the distributed Bragg reflectors (DBR) cavity simulations, the cell dimensions along the $x$ and $y$ directions were also fixed at 8 MEEP units, while the cell length along $z$-axis was adjusted according to the cavity resonance frequency.

Fig.~\ref{fig:DBR-cavity-geometry} demonstrates the geometry of the DBR cavity used in our simulations, which was formed by a pair of distributed Bragg mirrors separated by $\lambda$/2. The mirror layers were parallel to each other in the $x$--$y$ plane and were periodically arranged along the $z$ direction. Each side consisted of five layers of equidistant $\lambda$/8-thick dielectric media (refractive index $n$ =2, light blue rectangles). Adjacent dielectric layers were separated by air (refractive index $n$ =1) with a spacing of $\lambda$/4. The corresponding cavity frequency $\omega_\mathrm{c}$ was equal to 2$\pi$/$\lambda$.

For the NPoM nanocavity, simulations were performed using two-dimensional cylindrical coordinates. Fig.~\ref{fig:NPoM-geometry} depicts the geometry of the NPoM nanocavity used in this work. The nanocavity consisted of a 40 nm-diameter gold nanosphere, which was placed above a 70 nm-thick gold film. The gold nanosphere and the gold film were separated by a 1 nm-thick dielectric spacer with refractive index $n=1.4$. The dielectric response of both the gold nanosphere and gold film was described using the built-in Au material model in MEEP. 300 nm-thick air layers were included in two directions to allow sufficient propagation of the electromagnetic field. Finally, 300 nm-thick PMLs were also applied along the $r$ and $z$ boundaries to absorb outgoing radiation and suppress artificial reflections. 

For the NPoM nanocavity simulations, the MEEP length unit was chosen as $1 \mu\mathrm{m}$ to be consistent with the frequency-dependent dielectric parameters used for the built-in Au material model. To examine the numerical convergence of the NPoM nanocavity simulations, we first tested the dependence of the scattering spectrum of the bare cavity on the FDTD resolution and the thickness of air pad and PML. Fig.~\ref{fig:NPoM-spectra-convergence-test}(a) shows that the spectra obtained with 4000 and 5000 grid points per $\mu\mathrm{m}$ nearly overlap, indicating convergence of a resolution of 4000. We next varied the thickness of the air padding and PML regions from 200 to 400 nm while fixing the resolution at 4000. As shown in Fig.~\ref{fig:NPoM-spectra-convergence-test}(b), the peak position and overall shape of the scattering spectrum using 300 nm are close to those of 400 nm. We therefore used a resolution of 4000 and 300-nm-thick air padding and PML in the main text simulations.

We further examined the convergence of simulations for a single methylene blue molecule coupled to the NPoM nanocavity. Note that the lowest excitation energy of the methylene blue molecule at its optimized geometry, obtained from RT-TDDFTB, was 0.44 eV above the cavity frequency. To reduce the detuning, we selected a molecular geometry with an excitation energy of 1.88 eV from an NVT trajectory. Fig.~\ref{fig:molecule-NPoM-resolution-convergence}(a) shows that peak positions of lower and higher polaritons become stable when the FDTD resolution increases to 4000 and 5000. Similarly, energy evolution dynamics also converge at these resolutions (Fig.~\ref{fig:molecule-NPoM-resolution-convergence}(b)). These tests demonstrate that the numerical settings used in the main text are sufficient to obtain converged spectra and energy evolution dynamics of the coupled molecule--nanocavity system.

The electronic dynamics of Be, B$^+$ and C$^{2+}$ species and the ethylene molecule were simulated using the RT-TDDFT approach implemented in the CP2K software package \cite{hanasaki2025development}. We first applied a small perturbation (a $\delta$-kick of field strength of 0.001 a.u.) to the system along the $z$ direction at $t$=0. We let the system evolve 4000 simulation steps with a time step $\Delta t$=0.005 fs. The electronic excitation spectrum was then obtained through the Fourier transform of the dipole moment. After obtaining the first excitation energy, we performed RT-TDDFT simulations under an external excitation. The excitation was generated using the Gaussian source implemented in MEEP, whose temporal profile is described by:
\begin{equation*}
E_{\mathrm{ext}}(t)=E_0\exp\left[-\frac{(t-t_0)^2}{2\sigma^2}\right]
\cos\left[\omega_0(t-t_0)\right],
\end{equation*}
where $E_0$ is the field amplitude, $\sigma$ characterizes the temporal width of the Gaussian envelope, and $\omega_0$ is the central frequency consistent with atomic or molecular resonances. To obtain the energy evolution profile, we propagated 30000 steps with a time step $\Delta t$=0.005 fs.

All simulations were performed with a non-periodic $15\times15\times15~\text{\AA}^3$ unit cell. The PBE functional \cite{perdew1996generalized} was employed for all simulations. Kohn-Sham orbitals were expanded into a mixed Gaussian plane-wave basis, and the ionic cores were approximated by the Goedecker-Teter-Hutter (GTH) pseudopotentials \cite{goedecker1996separable}. Valence electrons were expanded with the DZVP-MOLOPT-SR-GTH basis set for all elements \cite{vandevondele2007gaussian}. Since atomic species and the ethylene molecule are closed-shell systems, all calculations were spin unpolarized.

The electronic dynamics of organic molecules in vacuum and cavities were simulated using RT-TDDFTB in DFTB+ \cite{Bonafe2020, Hourahine2020}. Electronic excitation spectra and energy profiles were obtained through similar procedures to those in Section~\ref{sec:RT-TDDFT}. The electronic structure was described at the DFTB3 level using the 3ob-3-1 Slater--Koster parameter set \cite{gaus2013parametrization}.

\section{Supplemental Figures}

\begin{figure}[H]
\centering
\includegraphics[width=0.5\textwidth]{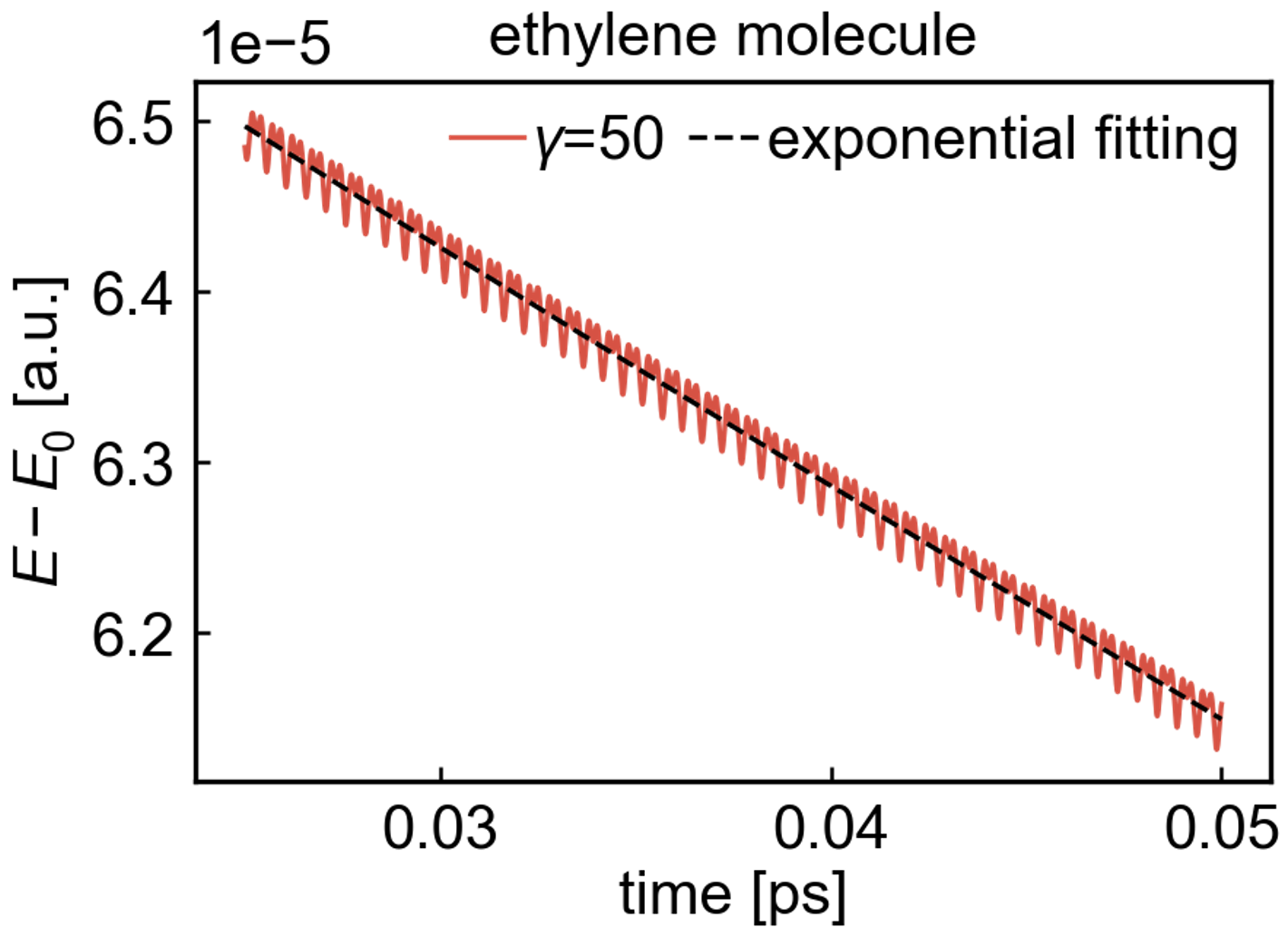}
\caption{Electronic energy (relative to the ground state) evolution for the ethylene molecule with rescaling factor $\gamma=50$. The decay rate obtained from exponential fitting is 2.20 ps$^{-1}$. The decay rate extrapolated to $\gamma=1$ is 0.881 ns$^{-1}$, corresponding to a lifetime of 1.13 ns.}
\label{fig:energy-evolution-ethylene}
\end{figure}

\begin{figure}[H]
\centering
\includegraphics[width=0.5\textwidth]{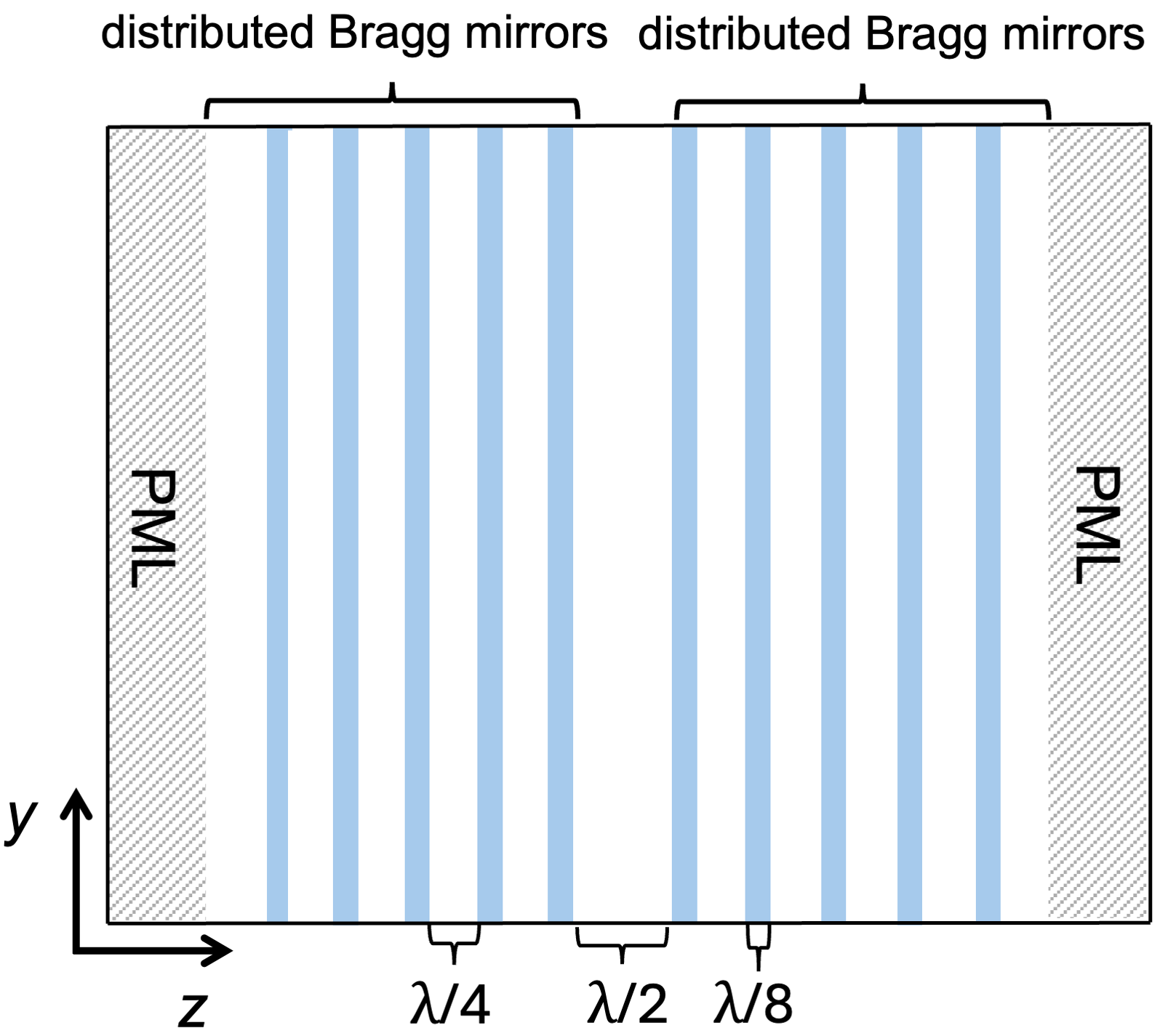}
\caption{Geometry of the DBR cavity used in simulations.}
\label{fig:DBR-cavity-geometry}
\end{figure}

\begin{figure}[H]
\centering
\includegraphics[width=0.7\textwidth]{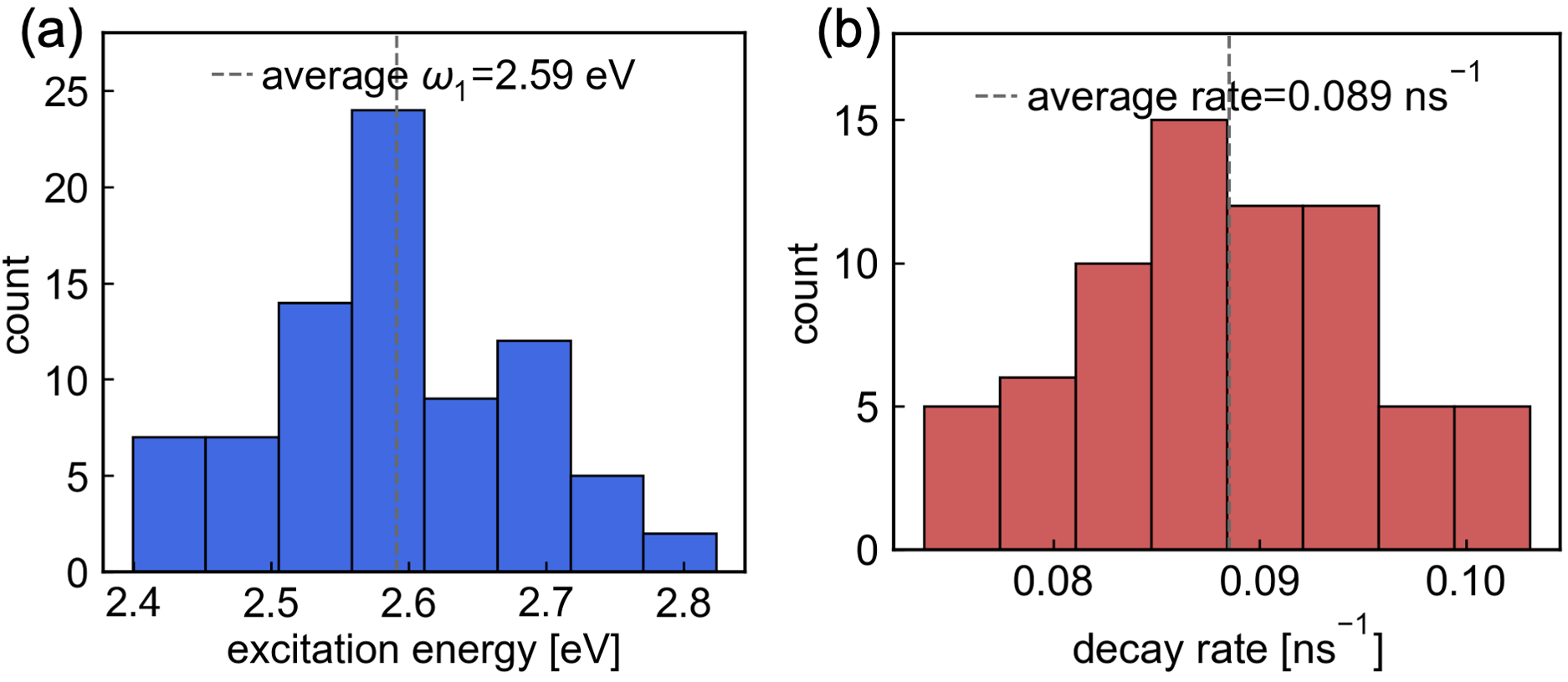}
\caption{Distributions of (a) excitation energies and (b) radiative decay rates in vacuum for 80 geometries selected from the NVT trajectory.}
\label{fig:excitation-energy-decay-rate-distribution}
\end{figure}

\begin{figure}[H]
\centering
\includegraphics[width=1.0\textwidth]{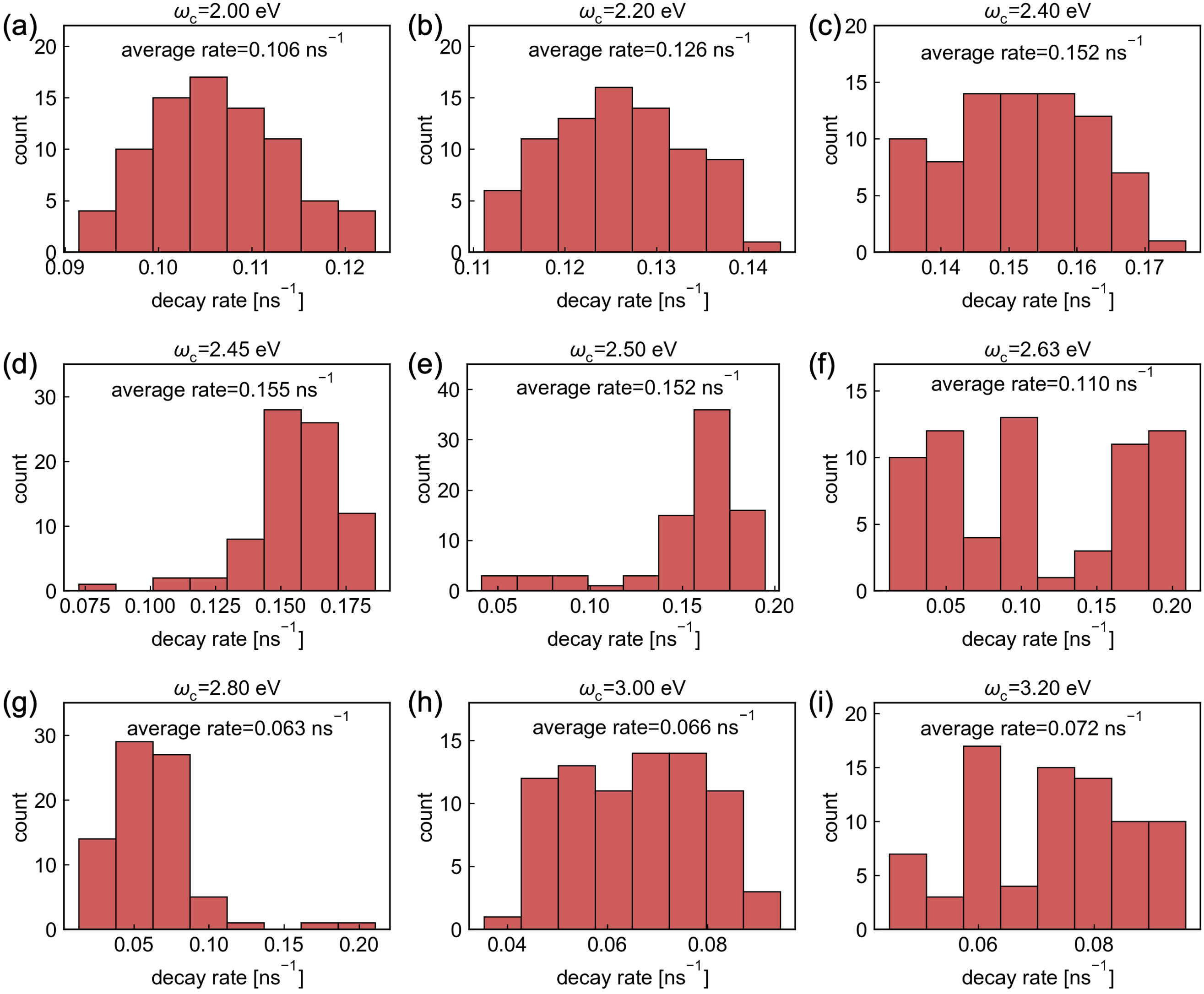}
\caption{Distributions of radiative decay rates for 80 geometries in the DBR cavity, with $\omega_\mathrm{c}=$ (a) 2.00, (b) 2.20, (c) 2.40, (d) 2.45, (e) 2.50, (f) 2.63, (g) 2.80, (h) 3.00 and (i) 3.20 eV.}
\label{fig:cavity-modified-decay-rate}
\end{figure}

\begin{figure}[H]
\centering
\includegraphics[width=0.5\textwidth]{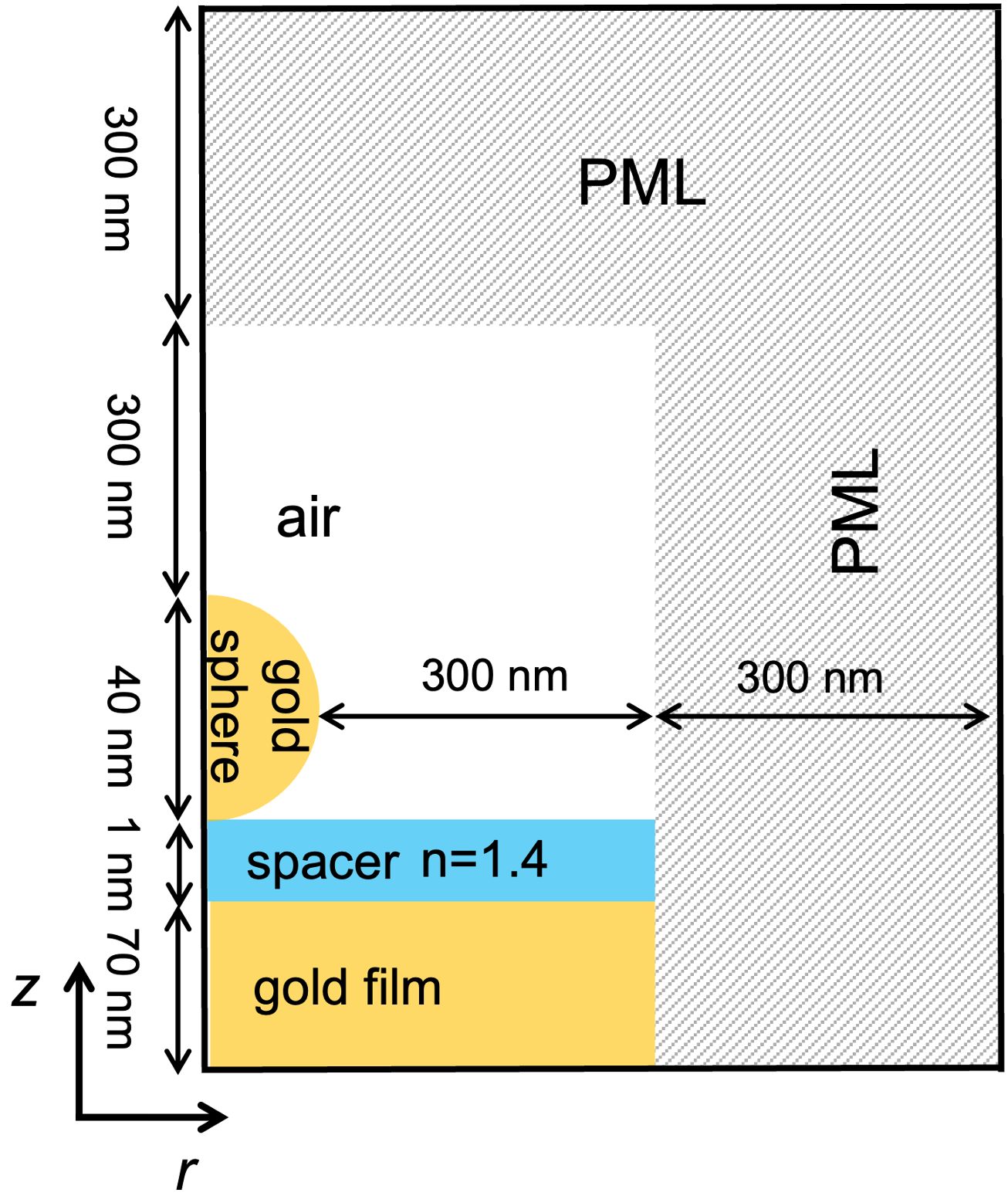}
\caption{Geometry of the simulated NPoM nanocavity in cylindrical coordinates.}
\label{fig:NPoM-geometry} 
\end{figure}

\begin{figure}[H]
\centering
\includegraphics[width=0.7\textwidth]{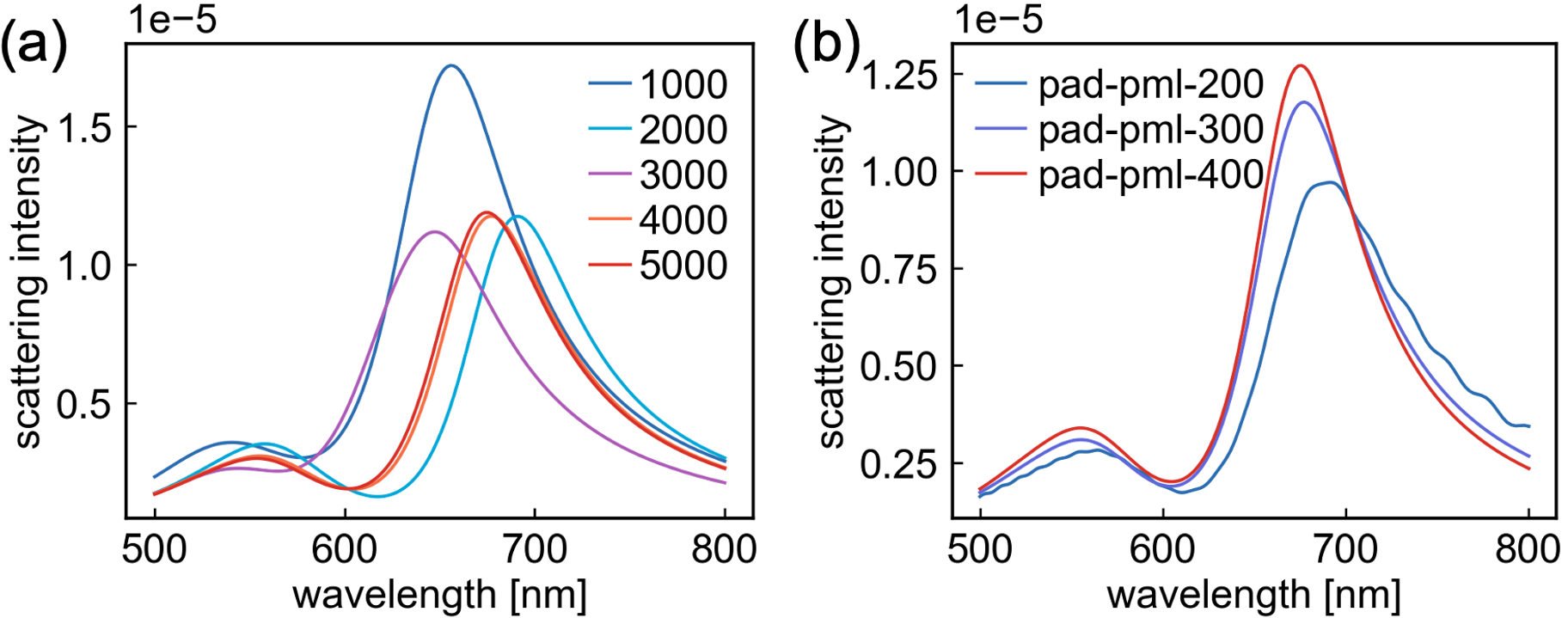}
\caption{(a) Scattering spectra of the NPoM nanocavity with the FDTD resolution ranging from 1000 to 5000 (the spatial grid spacing is reduced from 1.0 nm to 0.2 nm), accordingly, while the thicknesses of air pad and PML are fixed at 300 nm. (b) Scattering spectra of the NPoM nanocavity using 200, 300 and 400 nm-thick air pad and PML, with the FDTD resolution set to 4000.}
\label{fig:NPoM-spectra-convergence-test} 
\end{figure}

\begin{figure}[H]
\centering
\includegraphics[width=0.7\textwidth]{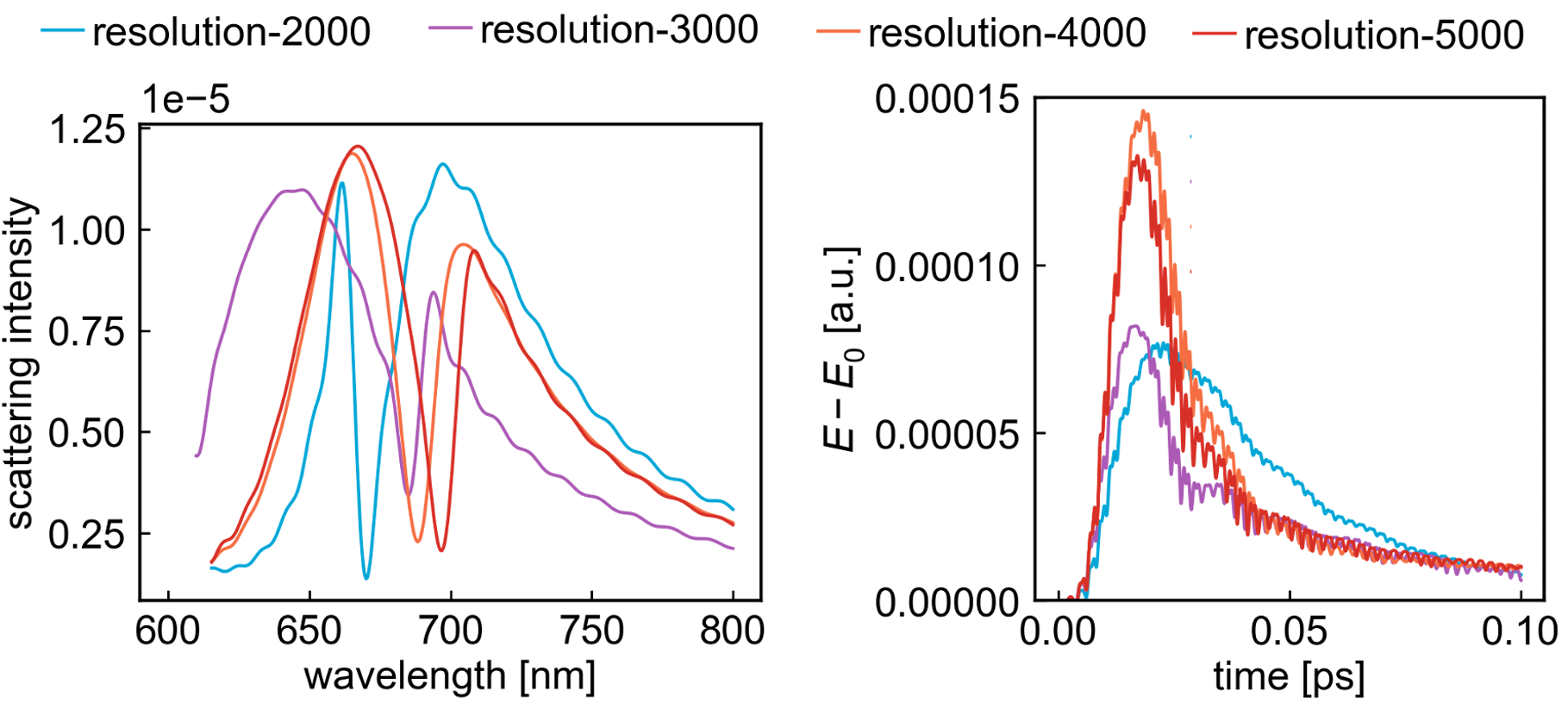}
\caption{(a) Scattering spectra of a methylene blue molecule in the NPoM nanocavity with different FDTD resolutions. (b) Energy evolution of a methylene blue molecule excited by a Gaussian pulse in the NPoM nanocavity with different resolutions.}
\label{fig:molecule-NPoM-resolution-convergence}
\end{figure}

\nocite{apsrev42Control}
\bibliographystyle{apsrev4-2}
\bibliography{aps_control, references_capitalized}